# Tritium and helium analyses in thin films by enhanced proton backscattering[*]

FU Tao(付涛) [1], AN Zhu(安竹) [1;1)], ZHU Jing-Jun(朱敬军) [1;1)], LIU Man-Tian(刘慢天) [1], MAO Li(毛莉)[2]

[1] Key Laboratory of Radiation Physics and Technology of Ministry of Education, Institute of Nuclear Science and Technology, Sichuan University, Chengdu 610064, China
[2] Institute of Nuclear Physics and Chemistry, China Academy of Engineering Physics, Mianyang, 621900, China

---

[*] Supported by Chinese National Fusion Project for ITER (2009GB106004), National Natural Science Foundation of China (11175123) and China Academy of Engineering Physics

1) Corresponding authors. E-mail: anzhu@scu.edu.cn, zhujingjun@scu.edu.cn

## ABSTRACT

In order to perform quantitative tritium and helium analysis in thin film sample by using enhanced proton backscattering (EPBS), EPBS spectra for several samples consisting of non-RBS light elements (i.e., T, $^4$He, $^{12}$C, $^{16}$O, $^{nat}$Si), medium and heavy elements have been measured and analyzed by using analytical SIMNRA and Monte Carlo-based CORTEO codes. The CORTEO code used in this paper is modified and some non-RBS cross sections of proton scattering from T, $^4$He, $^{12}$C, $^{14}$N, $^{16}$O and $^{nat}$Si elements taken from ENDF/B-VII.1 database and the calculations of SigmaCalc code are incorporated. All cross section data needed in CORTEO code over the entire proton incident energy-scattering angle plane are obtained by interpolation. It is quantitatively observed that the multiple and plural scattering effects have little impact on energy spectra for light elements like T, He, C, O and Si, and the RBS cross sections of light elements, instead of the non-RBS cross sections, can be used in SIMNRA code for dual scattering calculations for EPBS analysis. It is also observed that at the low energy part of energy spectrum the results given by CORTEO code are higher than the results of SIMNRA code and are in better agreement with the experimental data, especially when heavier elements exist in samples. For tritium analysis, the tritium depth distributions should not be simply adjusted to fit the experimental spectra when the multiple and plural scattering contributions are not completely accounted, or else inaccurate results may be obtained. For medium and heavy matrix elements, when full Monte Carlo RBS calculations are used in CORTEO code, the results from CORTEO code are in good agreement with the experimental results at the low energy part of energy spectra, at this moment quantitative tritium and helium analysis in thin film sample by using enhanced proton backscattering can be performed reliably.

*PACS number(s)*: 82.80.Yc, 25.40.Cm, 02.70.Uu

*Key words*: EPBS, Monte Carlo method, tritium analysis, multiple scattering, plural scattering

# 1 Introduction

Measurements of tritium and helium in materials play an important role in nuclear energy researches and in applications of nuclear technology, for example, in analyses for the first wall materials used in fusion reactor [1] and for tritium-containing targets used in neutron generator [2]. Among the analysis techniques developed for measuring tritium and helium in materials, ion beam analysis (IBA) techniques, including nuclear reaction analysis (NRA) [3,4], elastic recoil detection analysis (ERDA) [5,6] and enhanced proton backscattering (EPBS) [7,8] have been developed for many years and can provide information of tritium and helium concentration and depth distribution in materials in an almost nondestructive manner. In recent years, based on the work of Matsuyama, et al [9], we tried to develop the β-decay induced X-ray spectroscopy (BIXS) into a routine, accurate and *in situ* tritium analysis method for tritium-containing films by incorporating Monte Carlo simulation and Tikhonov regularization for dealing with the ill-posed inverse problems involved in the BIXS method [10-14]. We have employed the BIXS method to analyze tritium concentrations and depth distributions in tritium-containing Ti films with Mo substrate, and found that the total tritium concentrations obtained by the BIXS method were in good agreement with the results given by PVT method [13]. Meanwhile, we also carried out the EPBS analyses for tritium-containing Ti film samples and intended to examine whether the tritium depth distributions and concentrations given by the EPBS analysis are consistent with that given by the BIXS method. However, we found that, at the low-energy part where the signals from tritium appeared, the experimental EPBS spectra for the tritium-containing Ti film samples with Mo substrate can not be fitted well by using SIMNRA code [15], even including multiple and dual scattering in the fitting. Similar situation also occurred in EPBS analyses for helium-containing Ti film samples with Mo substrate. We examined several possible reasons for this situation [16], for example, the energy deposits in Au(Si) surface barrier detector used to detect backscattered protons due to tritium β-decay electrons and neutrons produced in the reaction $T(p,n)^3He$ when the incident energy of proton was larger than the threshold energy, 1.02 MeV, and possible inaccuracy of non-Rutherford backscattering (non-RBS) cross sections of T(p,p)T. We observed that these possible reasons can not explain the disagreement between the experimental EPBS spectra and the fitting spectra given by SIMNRA code at the low-energy part for tritium analysis. Moreover, we noticed that SIMNRA code utilizes Rutherford backscattering (RBS) cross sections, instead of non-RBS cross sections which are usually one to three orders of magnitude larger than RBS cross sections ( in particular for tritium, the ratio of non-RBS cross sections to RBS cross sections can be ~1000 at ~3.5 MeV [7]), to calculate the dual scattering contributions for non-RBS light elements (e.g., T, $^4$He, $^{12}$C, $^{14}$N, $^{16}$O, $^{nat}$Si and so on). For small-angle scattering, non-RBS cross sections tend to be equal to RBS cross sections, therefore whether non-RBS cross sections or RBS cross sections are used are not important in this case. However, for large-angle plural scattering, non-RBS cross sections possibly play an important role, and SIMNRA code (version 6.06, the newest version now) also warns users that inaccurate results will possibly be given when EPBS spectra for non-RBS light elements are analyzed by using RBS cross sections for calculating the dual scattering contributions. Therefore, whether or not RBS cross sections, instead of non-RBS cross sections, can be used to calculate the large-angle plural scattering contributions for non-RBS light elements still needs quantitative verification. On the other hand, SIMNRA code also neglects the higher order large-angle scattering contributions with more than two scattering events. Barradas [17] also pointed out some causes of RBS analytical model (as opposed to Monte Carlo method) for the disagreement between experiments and calculation results from analytical models at the low-energy part of RBS spectra and developed an approximate analytical method to deal with this issue. Because Monte Carlo methods can result in a very realistic simulation of RBS spectrum and quantitative analyses for tritium and helium in thin films are needed by us, therefore, in this paper, we will employ Monte Carlo method for EPBS spectrum analysis (i.e., CORTEO code [18]), based on the non-RBS cross sections of proton backscattering from tritium and helium, to examine whether the disagreement between the experimental EPBS spectra and the fitting spectra given by SIMNRA code at the low-energy part can be quantitatively and accurately explained. In addition, in order to further investigate the effect of RBS cross sections, instead of non-RBS cross sections, being used to calculate the dual scattering contributions by SIMNRA code when non-RBS elements exist in samples, we also analyze some other experimental EPBS spectra of samples which consist of non-RBS elements, e.g., $SiO_2$ and SiC. Because tungsten is the most important candidate for the first wall materials of fusion reactor, and retention of tritium and helium will be possibly happened when tungsten is used as the first wall materials in fusion reactor and therefore need to be analyzed, we also analyze the experimental EPBS spectrum of tungsten in this paper.

This paper is organized as follows. Section 2 describes the sample preparation and EPBS experiment, Section 3 introduces the codes used in this paper and code modification. Section 4 shows the results and discussion. The conclusions are given in Section 5.

## 2 Experimental

### 2.1 sample preparation

The tritium-containing Ti film sample was prepared by firstly evaporating Ti onto a smooth Mo substrate, and then was placed in a tritium gas to absorb tritium [13]. The tritium content absorbed in the Ti film sample was determined by the pressure change of tritium gas based on the equation of state of ideal gas, i.e., PVT method. The thickness of Ti film of this sample was about 5 μm, the thickness of Mo substrate was about 1 mm, T/Ti ratio in the Ti film measured by the PVT method was about 1.51. The tritium depth distribution in this sample prepared by the processes described above usually should be uniform or the tritium concentration decreases as the depth increases. The helium-containing Ti film sample was fabricated by depositing Ti onto smooth Mo or Si substrate using magnetron co-sputtering in a gas mixture of argon and helium, the details for sample fabrication was given in Ref. [19]. The thickness of Ti film of this sample was about 1.5 μm, the thickness of Mo or Si substrate was about 1 mm, He/Ti ratio in the Ti film measured by the EPBS method was about 0.60. The helium depth distribution in this sample usually should be uniform [19]. $SiO_2$, SiC and W samples were of high-purity (>99.9%) and about 1 mm thick, the sample surfaces were polished.

## 2.2 EPBS experiments

The EPBS experiments were performed at the 2.5 MeV Van de Graaff accelerator at the Institute of Nuclear Science and Technology of Sichuan University. The incident proton energy was about 2 MeV, the beam spot size was about 2 mm of diameter, the direction of incident proton beam was vertical to the sample surface. A semiconductor Au(Si) surface barrier detector with a depletion depth of 100 μm was placed in the target chamber at the angle of $160^{o}$ or $165^{o}$ with respect to the incident proton beam direction to record the backscattered protons. The intensity of incident proton beam was adjusted to keep the dead time correction less than 2%.

## 3 Codes and modifications

Several software packages, analytical or Monte Carlo-based, have been developed for many years to perform NRA, ERDA, RBS and EPBS analyses, the status of these codes was reviewed in Ref. [20]. The comparisons among some codes have also been extensively made with respect to many aspects [21], and their advantages and weaknesses have been discussed [22]. In this paper, analytical SIMNRA code and Monte Carlo-based CORTEO code are utilized.

### 3.1 SIMNRA code

SIMNRA code [15] is a widely-used Microsoft Windows program with full graphical user interface for the simulation of back or forward scattering spectra for IBA techniques, e.g., NRA, ERDA, RBS and non-RBS. About several hundred different non-Rutherford and nuclear reaction cross sections for incident protons, deuterons, He-ions are included. New cross section data can be added by users in R33 file format, for example, from IBANDL database [23] or from theoretical calculations by SigmaCalc code developed by Gurbich [24,25]. Correction factors by L'Ecuyer or by Andersen can be applied in SIMNRA code due to partial screening of nuclear charges by the electron shells surrounding nuclei. SIMNRA code can use several different sets of stopping power data for the stopping of light and heavy ions in all elements, e.g., Andersen-Ziegler stopping, Ziegler-Biersack stopping, KKK stopping and SRIM stopping, and it can also use stopping power data that users defined. Bohr's model, Chu's model or Yang's model for energy loss straggling can be used in SIMNRA code. This code can treat the surface roughness of sample for two cases, i.e., rough film on a smooth substrate and smooth film on a rough substrate. SIMNRA code can also take into account multiple (small angle) scattering and dual (large angle) scattering as an approximate calculation of plural scattering. However, SIMNRA code uses Rutherford scattering cross sections for the dual scattering calculation when non-RBS elements exist in samples, this may result in inaccurate results.

3.2 CORTEO code

CORTEO code is a Monte Carlo-based program that simulates ion beam analysis spectra, i.e., RBS and ERDA, and is freely available with its source code under the terms of the GNU General Public License. By simulating the trajectory of each ion, it can take into account more naturally and accurately some effects such as multiple and plural scattering. Some improvements have been made so that the simulations can be achieved with sufficient statistics on a personal computer in a reasonable amount of time [18]. Correction factors by Andersen for RBS cross sections due to partial screening of nuclear charges by the electron shells can be used in CORTEO code. Stopping power data are obtained from SRIM's SRModule. Bohr's model, Chu's model or Yang's model for energy loss straggling can be used in this code.

Monte Carlo analysis for EPBS spectra needs non-RBS cross sections of proton scattering over the entire proton incident energy-scattering angle plane in question. Therefore, for the purpose of this paper, we have modified CORTEO code and incorporated some non-RBS cross sections of proton scattering from T, $^4$He, $^{12}$C, $^{14}$N, $^{16}$O and $^{nat}$Si elements into this code. The cross sections of EPBS for tritium are taken from ENDF/B-VII.1 database, which are based on R-matrix analysis [26]. The cross sections of EPBS for $^4$He, $^{12}$C, $^{14}$N, $^{16}$O and $^{nat}$Si elements are taken from the calculations of SigmaCalc code [24,25]. These cross sections are taken at certain grids of incident proton energies and scattering angles. All cross sections needed in the Monte Carlo analysis are obtained by interpolation. In particular, the ratios of these non-RBS to corresponding RBS cross sections at the scattering angle of $0^o$ or below the energies determined by Bozoian's formulae [27] are set to be one. The cross sections data at $165^o$ for tritium from ENDF/B-VII.1 database are also compared with the available experimental data [7], and they are in good agreement [16].

**4 Results and discussion**

In this section, we utilize SIMNRA and modified CORTEO codes to analyze the experimental EPBS spectra we obtained. For being comparable, the calculations based on these two codes are performed under the same conditions, i.e., experimental setup and target structure used in these two codes are same, and correction factor by Andersen, SRIM stopping power and Yang's model for energy loss straggling are used in these two codes, and the EPBS cross sections described in Section 3.2 are also used in these two codes. The cone angles used in CORTEO code are determined according to the method given in its users' manual [18]. In addition, the simulated spectra given by SIMNRA code in the following discussion are the results calculated with multiple and dual scattering and the default cutoff energy (i.e., 10 keV). SIMNRA code can only use Rutherford backscattering (RBS) cross sections, instead of non-RBS cross sections, to calculate the dual scattering contributions for non-RBS light elements.

Fig.1 and Fig.2 show the measured and simulated spectra for 2 MeV proton backscattered from $SiO_2$ and SiC samples at a scattering angle of $160^o$. We have compared simulated spectra given by SIMNRA code with and without multiple and dual scattering although the dual scattering contributions are calculated with RBS cross sections instead of non-RBS cross sections, and found that they are almost the same. As can be seen for $SiO_2$ sample in Fig.1, the result calculated with SIMNRA code overall is in good agreement with the experimental data and the result given by CORTEO code except at the very low energy part where the result given by CORTEO code is higher than the result of SIMNRA code and is closer to the experimental data. For SiC sample, in Fig.2 the result calculated with SIMNRA code also overall is in good agreement with the experimental data and the result given by CORTEO code except around the carbon resonance peak where the results given by SIMNRA and CORTEO codes are higher than the experimental data. We think that this may be caused by simulating without considering the surface roughness and the possible inaccuracy of cross sections used here around the carbon resonance. On the other hand, the results given by these two codes are also somewhat different around the carbon resonance, this difference may be due to the different treatments of the cross section and straggling, which have been discussed in detail in Section 3.4 of Ref. [21]. The difference between the experiment and the simulations at the very low energy part may be due to SIMNRA code's neglect of the higher order large-angle scattering contributions with more than two scattering events whereas CORTEO code taking into account all scattering events.

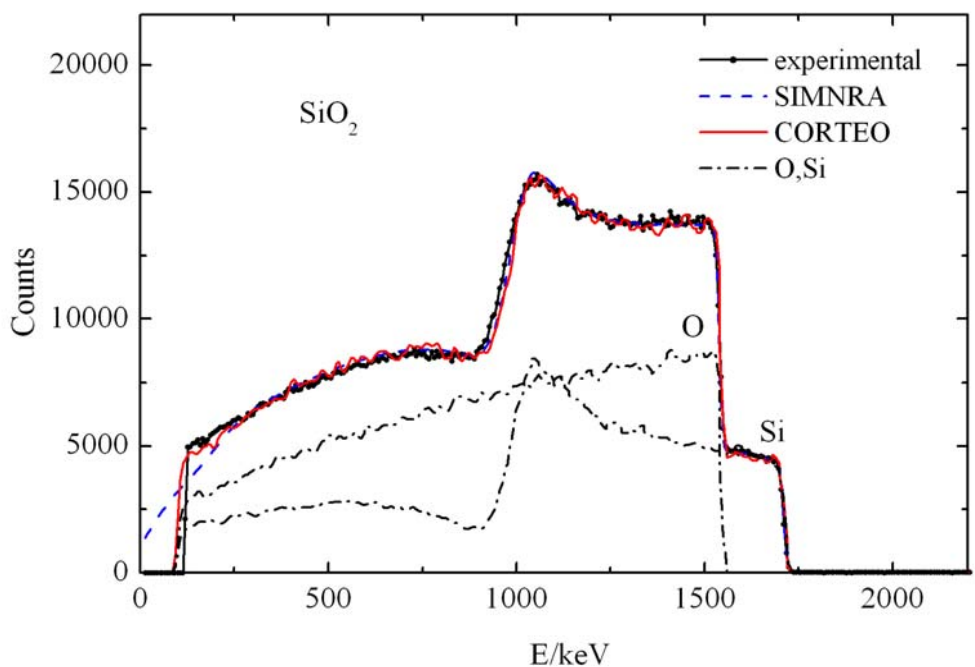

**Fig. 1.** (color online) Comparison of the experimental and simulated energy spectra with SIMNRA and CORTEO codes for 2 MeV proton backscattered from thick $SiO_2$ sample at a scattering angle of $160^o$. The simulated spectrum given by SIMNRA code is calculated with multiple and dual scattering. Individual elemental spectra calculated from CORTEO code are also shown.

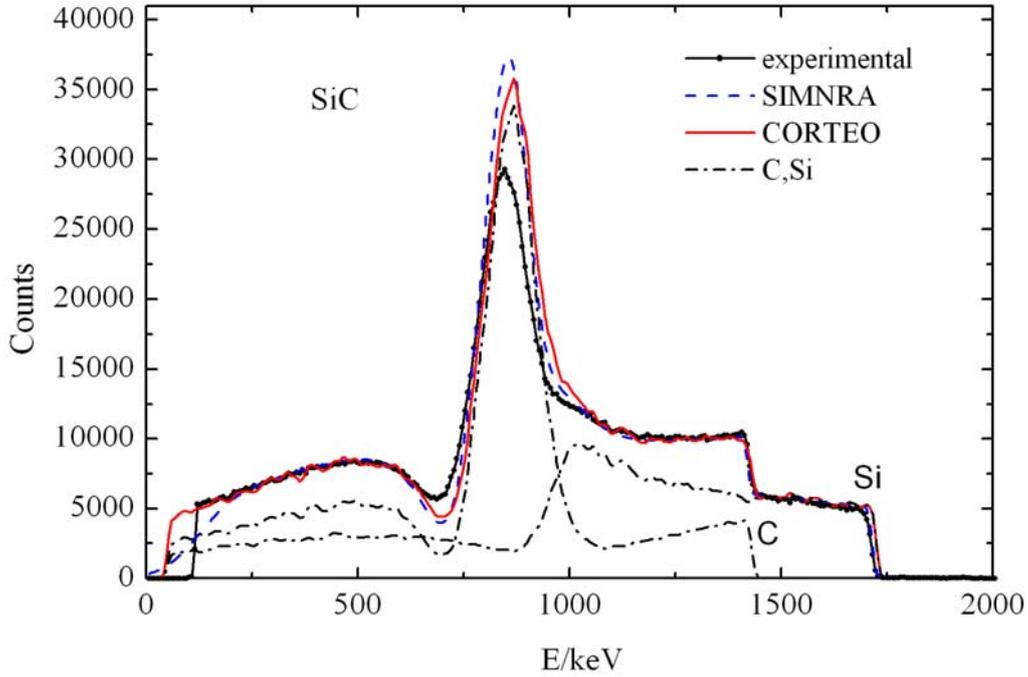

**Fig. 2.** (color online) The same as Fig.1 but for 2 MeV proton backscattered from the thick SiC sample.

The results above also indicate that multiple and plural scattering effects have little impact on energy spectrum for light elements like C, O and Si except at very low energy part, therefore RBS cross sections for EPBS analysis of light elements, instead of non-RBS cross sections, used in SIMNRA code for dual scattering calculations are acceptable. This result can be approximately understood as follows: when a proton collides with a light nucleus, the backscattered proton will have a larger energy loss (in comparison with the case of a proton colliding with a heavier nucleus) and the energy of the backscattered proton decreases rapidly, therefore, the non-RBS cross sections for the backscattered proton approach to the RBS cross sections. Moreover, because the RBS cross sections are proportional to $Z^2$ ($Z$ is the atomic number of target atom), the probability for plural scattering becomes very small for light elements even the non-RBS cross sections are large.

Fig.3 and Fig.4 show the measured and simulated spectra for 2 MeV proton backscattered from the helium-containing Ti film samples with smooth Si and Mo substrates at a scattering angle of $160^{\circ}$. The RBS cross sections for He and Si elements are used in SIMNRA code for dual scattering calculations. The helium depth distributions are reasonably assumed to be uniform. From Fig.3 and Fig4, we can see that the results obtained from SIMNRA code overall are in good agreement with the experimental data and the results given by CORTEO code except at the low energy part where the results given by CORTEO code are higher than the results of SIMNRA code and are closer to the experimental data. We notice that the differences among the results of SIMNRA and CORTEO codes and experimental data are originated from heavier substrate elements, e.g., Mo, and the heavier the substrate elements are, the larger the differences are. This difference may also be due to CORTEO code taking into account all scattering events. However, for this two samples, the helium analyses have not yet been affected. In addition, the results shown in Fig.3 and Fig.4 also indicate, as in Fig.1 and Fig.2, that RBS cross sections for EPBS analysis of light elements, instead of non-RBS cross sections, can be used in SIMNRA code for dual scattering calculations.

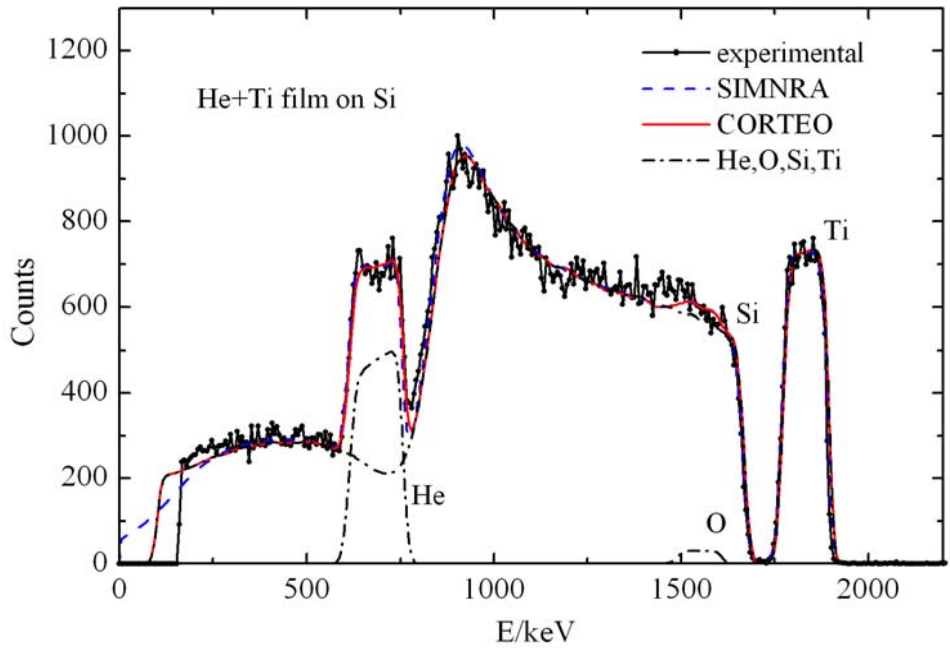

**Fig. 3.** (color online) The same as Fig.1 but for 2 MeV proton backscattered from the He-containing Ti film sample with smooth thick Si substrate.

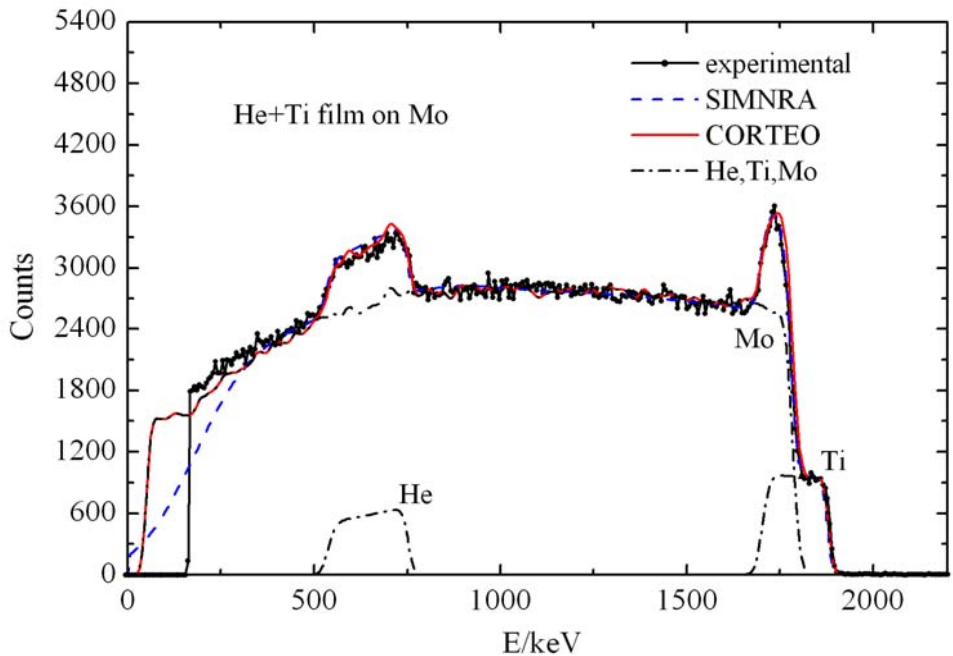

**Fig. 4.** (color online) The same as Fig.1 but for 2 MeV proton backscattered from the He-containing Ti film sample with smooth thick Mo substrate.

In Figs.5, the measured and simulated spectra for 2 MeV proton backscattered from the tritium-containing Ti film sample with smooth Mo substrate at a scattering angle of $165^{o}$ are presented. The RBS cross sections for T element are used in SIMNRA code for dual scattering calculations. The tritium depth distribution is reasonably assumed to be uniform. From Fig.5, we can observe that the result obtained from SIMNRA code overall is in good agreement with the experimental data and the result given by CORTEO code except around the low energy spectrum where the signals from tritium appear, at which the result given by CORTEO code is higher than the result of SIMNRA code and is closer to the experimental data. This comparison indicates that the difference between the result obtained from SIMNRA code and experimental data should not be solved simply by adjusting the tritium depth distribution, or else inaccurate results for tritium analysis may be obtained. In Fig.5, we also show the comparison of individual spectra of tritium from CORTEO code and from SIMNRA code with single scattering model, we can see that they are almost same, and this further indicates that the multiple and plural scattering contributions from light elements, e.g., tritium, are not important even the non-RBS cross sections are very larger than corresponding RBS cross sections and the differences among the results of SIMNRA and CORTEO codes and experimental data are mainly due to the multiple and plural scattering contributions from the heavier substrate element, e.g., Mo.

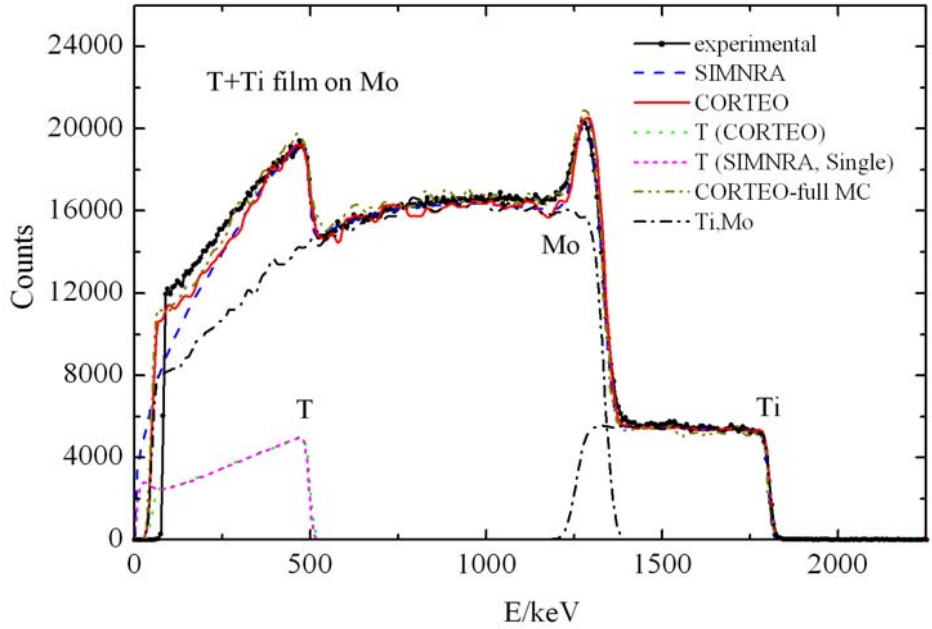

**Fig. 5.** (color online) Comparison of the experimental and simulated energy spectra with SIMNRA and CORTEO codes for 2 MeV proton backscattered from the T-containing Ti film sample with smooth thick Mo substrate at a scattering angle of 165°. The total simulated spectrum given by SIMNRA code is calculated with multiple and dual scattering, the individual spectra of tritium from CORTEO code and from SIMNRA code with single scattering model are shown, and other individual elemental spectra calculated from CORTEO code are also shown. A full Monte Carlo calculation is also shown, which is the sum of full RBS calculations for Ti and Mo and RBS calculation for T by CORTEO code.

Finally, the measured and simulated spectra for 2 MeV proton backscattered from W are shown in Fig.6. We can see that at the low energy part of the spectrum both CORTEO and SIMNRA codes can not give satisfactory agreement with the experimental result. In fact, in Fig.5, although the result from CORTEO code is better than that from SIMNRA code at the low energy part when comparing with the experimental data, the result from CORTEO code also needs to be improved. This disagreement may be caused by approximate algorithms in codes. For example, although in CORTEO code some improvements have been made in order that the computing time can be decreased by several orders of magnitude, at the same time these improvements also introduce some problems, which have been pointed out in Ref.[22]. Therefore, in Figs.5-6, we also show the results of full Monte Carlo RBS calculations (by using CORTEO code on a computer cluster) for Ti, Mo and W. We can see that the full Monte Carlo results improve the agreement with the experimental spectra at the low energy part, and hence under these circumstances where the multiple and plural scattering contributions are sufficiently accounted the quantitative tritium and helium analysis in thin film sample by using enhanced proton backscattering can be performed reliably. Although a full Monte Carlo calculation requires long computing times for practical applications, an analytical model simulation can be first performed and then followed by a full Monte Carlo calculation when necessary, for example, for our cases of quantitative tritium and helium analyses. In addition, some other causes, for example, slit scattering, low-energy component in the beam, inaccurate physical data (cross section, stopping power and so on) or unaccounted physical phenomenon, may also contribute to the disagreement between experiments and calculations at the low energy part of RBS spectra, some of which have been discussed in Refs.[21,22,28,29]. For our cases, we observe that different stopping powers, provided by SIMNRA code, can lead to apparent differences for thick W target at the low energy part of EPBS spectrum, while for thick Mo target the calculation results are relatively stable and the differences are smaller for different stopping powers.

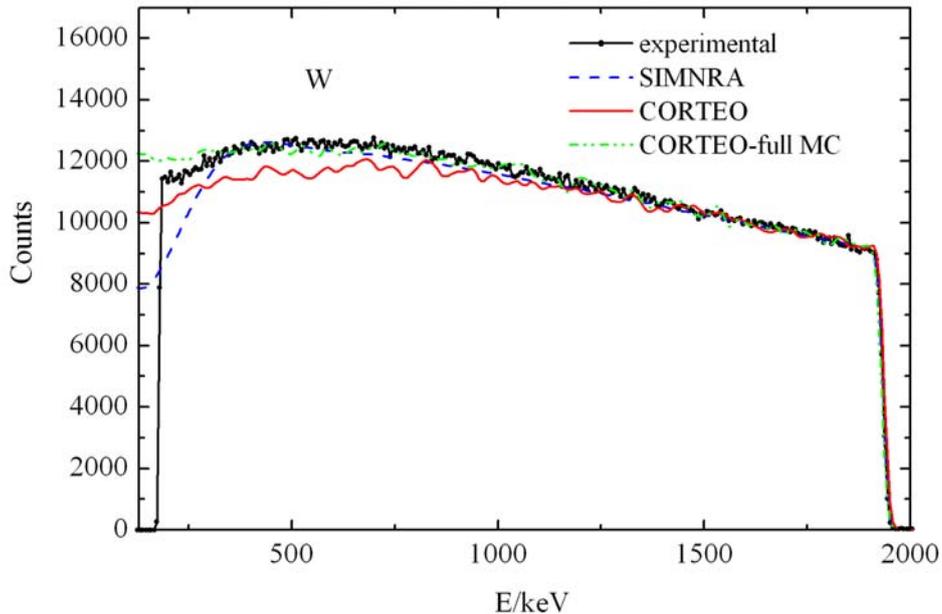

**Fig. 6.** (color online) The same as Fig.1 but for 2 MeV proton backscattered from smooth thick W sample. A full Monte Carlo calculation by CORTEO code is also shown.

# 5 Conclusions

EPBS spectra for several samples consisting of non-RBS light elements, medium and heavy elements have been measured and analyzed by using analytical SIMNRA and Monte Carlo-based CORTEO codes. The CORTEO code is modified and some non-RBS cross sections of proton scattering from T, $^4$He, $^{12}$C, $^{14}$N, $^{16}$O and $^{nat}$Si elements are incorporated. We quantitatively observe that the multiple and plural scattering effects have little impact on energy spectra for light elements like T, He, C, O and Si, and the RBS cross sections of light elements, instead of the non-RBS cross sections even they are very larger than corresponding RBS cross sections, can be used in SIMNRA code for dual scattering calculations for EPBS analysis. We also observe that at the low energy part the results given by CORTEO code are higher than the results of SIMNRA code and are closer to the experimental data, especially when heavier elements exist in samples. This may be caused by SIMNRA code neglecting the higher order large-angle scattering contributions with more than two scattering events whereas CORTEO code taking into account all scattering events. For tritium analysis, the tritium depth distributions should not be simply adjusted to fit the experimental spectra when the multiple and plural scattering contributions are not completely accounted, or else inaccurate results may be obtained. For medium and heavy matrix elements, when full Monte Carlo RBS calculations are used in CORTEO code, the results from CORTEO code are in good agreement with the experimental results at the low energy part of EPBS spectra, at this moment quantitative tritium and helium analysis in thin film sample by using enhanced proton backscattering can be performed reliably. For practical applications, an analytical model simulation can be first performed and then followed by a (full) Monte Carlo calculation when necessary.

## Acknowledgements

The authors gratefully acknowledge the crew of the Van de Graaff accelerator at Institute of Nuclear Science and Technology, Sichuan University for accelerator operation during this study.